\documentclass[a4paper,german,12pt]{article}
\usepackage{graphicx}
\usepackage{txfonts}
\usepackage{color}
\usepackage{natbib}
\usepackage{setspace}
\usepackage{lineno}
\bibpunct{(}{)}{;}{a}{}{,} 

\doublespace

\begin{document}


\textbf{The dependence of the ice-albedo feedback on atmospheric properties}
\newline
\newline

P. von Paris$^{1,2,3}$, F. Selsis$^{1,2}$, D. Kitzmann$^{4}$, H. Rauer$^{3,4}$
\newline
\newline

$^1$ Univ. Bordeaux, LAB, UMR 5804, F-33270, Floirac, France

$^2$ CNRS, LAB, UMR 5804, F-33270, Floirac, France

$^3$ Institut f\"{u}r Planetenforschung, Deutsches Zentrum f\"{u}r Luft- und Raumfahrt, Rutherfordstr. 2, 12489 Berlin, Germany

$^4$ Zentrum f\"{u}r Astronomie und Astrophysik, Technische Universit\"{a}t Berlin, Hardenbergstr. 36, 10623 Berlin, Germany
\newline
\newline
\newline
\newline

Corresponding author:

Philip von Paris

Email: philip.vonparis@dlr.de

Tel.: +49 (0) 30- 67055 7939

\newpage

\section*{Abstract}

The ice-albedo feedback is a potentially important de-stabilizing effect for the climate of terrestrial planets. It is based on the positive feedback between decreasing surface temperatures, an increase of snow and ice cover and an associated increase in  planetary albedo, which then further decreases surface temperature. A recent study shows that for M stars, the strength of the ice-albedo feedback is reduced due to the strong spectral dependence of stellar radiation and snow/ice albedos, i.e. M stars primarily emit in the near-IR, where the snow and ice albedo is low, and less in the visible where the snow/ice albedo is high.

This study investigates the influence of the atmosphere (in terms of surface pressure and atmospheric composition) for this feedback, since an atmosphere was neglected in previous studies. A plane-parallel radiative transfer model is used for the calculation of planetary albedos. We varied CO$_{2}$ partial pressures as well as the H$_2$O, CH$_4$, and O$_3$ content in the atmosphere for planets orbiting Sun-like and M-type stars.

Results suggest that for planets around M stars, the ice-albedo effect is significantly reduced, compared to planets around Sun-like stars. Including the effects of an atmosphere further suppresses the sensitivity to the ice-albedo effect. Atmospheric key properties such as surface pressure, but also the abundance of radiative trace gases can considerably change the strength of the ice-albedo feedback. For dense CO$_2$ atmospheres of the order of a few to tens of bar, atmospheric rather than surface properties begin to dominate the planetary radiation budget. At high CO$_2$ pressures, the ice-albedo feedback is strongly reduced for planets around M stars. The presence of trace amounts of H$_2$O and CH$_4$ in the atmosphere also weakens the ice-albedo effect for both stellar types considered. For planets around Sun-like stars, O$_3$ could also lead to a very strong decrease of the ice-albedo feedback at high CO$_2$ pressures.
\newpage

\section{Introduction}

The ice-albedo feedback is a very important positive feedback for the Earth's climate. In simple words, it describes the possible runaway cooling of the planetary surface. Decreasing surface temperature leads to an increase of snow and ice cover, which increases the surface albedo. Hence, more incoming stellar radiation is reflected back to space, which in turn decreases surface temperature.

A recent work by \citet{joshi2012} connected this ice-albedo feedback to exoplanets and the habitable zone (HZ). The HZ is defined as the region around a star where a planet with a suitable atmosphere could have liquid water on the surface, hence the potential for life as we know it \citep{kasting1993}. Calculations presented by \citet{joshi2012} showed that the planetary albedos for planets orbiting M stars are much lower than for planets orbiting Sun-like stars because of the strong spectral dependence of both ice and snow albedo and the stellar spectrum. Hence, the ice-albedo feedback is much less important for planets orbiting M stars. Based on these results, \citet{joshi2012} stated that the outer boundary of the HZ is more extended around M stars than around G stars since planetary albedos are much lower.

However, the ice-albedo feedback is not the key factor when determining the extent of the outer boundary of the HZ. It is usually assumed that geochemical cycles such as the carbonate-silicate cycle (e.g., \citealp{walker1981,abbot2012}) will stabilize the climate (e.g., \citealp{kasting1993,selsis2007gliese}). The outer boundary of the HZ is then determined by the interplay between an increase of planetary albedo with increasing CO$_2$ and the atmospheric greenhouse effect provided by CO$_2$ \citep{kasting1993}. A higher planetary albedo does not necessarily imply lower surface temperatures, as has been shown by, e.g., \citet{mckay1989titan}, or \citet{mischna2000}. Still, at one point, an increase in planetary albedo will lead to cooler surface temperatures, an effect known for CO$_2$ as the maximum greenhouse effect \citep{kasting1993}. Furthermore, atmospheres towards the outer edge of the HZ are assumed to be very dense (of the order of several bars of CO$_2$, \citealp{kasting1993}), implying that the influence of surface albedo on the planetary energy budget is probably rather small (see, e.g., \citealp{wordsworth2011} for 3D investigations of the outer HZ). Therefore, conclusions on habitability and the extent of the HZ cannot be easily drawn from the investigation of the ice-albedo feedback.

Nevertheless, the ice-albedo feedback is still an important process for the atmospheric and climate evolution of the Earth. It is generally accepted that Earth encountered several global glaciation episodes (the so-called snowball Earth, e.g., \citealp{kirschvink1992}, \citealp{hoffman1998}, \citealp{hyde2000}, \citealp{kastono2006}) and, more recently, ice ages with widespread glaciations down to mid-latitudes (e.g., \citealp{tarasov1997}). Some of these snowball episodes coincide with major events in the evolution of the biosphere, such as the advent of atmospheric O$_2$ \citep{kasthoward2006,goldblatt2006} and multicellular organisms \citep{hyde2000}. Rapid, perhaps geologically or biologically induced, changes in atmospheric composition might have been responsible for triggering these glaciations (e.g., \citealp{pavlov2003}).

Therefore, it is of interest for astrobiology and a comprehensive theory of atmospheric evolution to investigate whether planets orbiting other central stars might also be subject to such feedbacks. The work by \citet{joshi2012} calculated planetary albedos based on surface albedos and did not take explicitly the atmosphere into account. However, atmospheric properties such as surface pressure or atmospheric composition have a potentially large influence on the planetary albedo, hence the ice-albedo feedback. In this work, we present similar calculations as in \citet{joshi2012}, but with a plane-parallel radiative transfer model which incorporates the radiative effects of an atmosphere. The model is used to calculate planetary albedos of prescribed atmospheric scenarios when varying atmospheric properties (such as atmospheric composition and surface pressure). In a complementary approach, \citet{shields2013} used a combination of 2D and 3D models to assess the development of ice and snow cover on planets orbiting M, G, and F stars upon decreasing stellar insolation. However, the considered planetary atmospheres were Earth-like in terms of surface pressure and composition. With a 1D radiative transfer model, \citet{shields2013} also calculated planetary albedos as a function of CO$_2$ partial pressure, however without considering additional trace species, such as water, ozone, or methane.

The paper is organized as follows: Sect. \ref{model} presents the model and scenarios, Sect. \ref{results} the results and Sect. \ref{discuss} a discussion of additional aspects. We summarize the results and conclude with Sect. \ref{summary}.

\section{Methods}

\label{model}

\subsection{Radiative transfer}

\label{radtra}

We used a plane-parallel radiative transfer model to obtain planetary albedos of a wide range of prescribed, fixed atmospheric scenarios. The radiative fluxes important for the planetary albedo (i.e., stellar radiation) are calculated for a spectral range of 0.237-4.545\,$\mu$m, divided into 38 spectral intervals. Scattering is treated based on a $\delta$-Eddington-2-stream algorithm summarized by \citet{toon1989}. The frequency integration is performed using correlated-k coefficients in up to 4-term exponential sums \citep{wiscombe1977}. Gaseous opacities are taken from \citet{pavlov2000} and \citet{karkoschka1994}. Rayleigh scattering is incorporated using approximations presented by \citet{vparis2010gliese}, \citet{vardavas1984}, and \citet{allen1973}.

The pressure grid in the model atmospheres is determined from the surface pressure $p_{\rm{surf}}$ up to a pressure of 6.6$\times$10$^{-5}$ bar at the model lid, divided into 52 model layers, approximately spaced equidistantly in $\log$\,(pressure),

\subsection{Model procedure}

\label{procedure}

\begin{figure}[h]
    \resizebox{\hsize}{!}{\includegraphics*{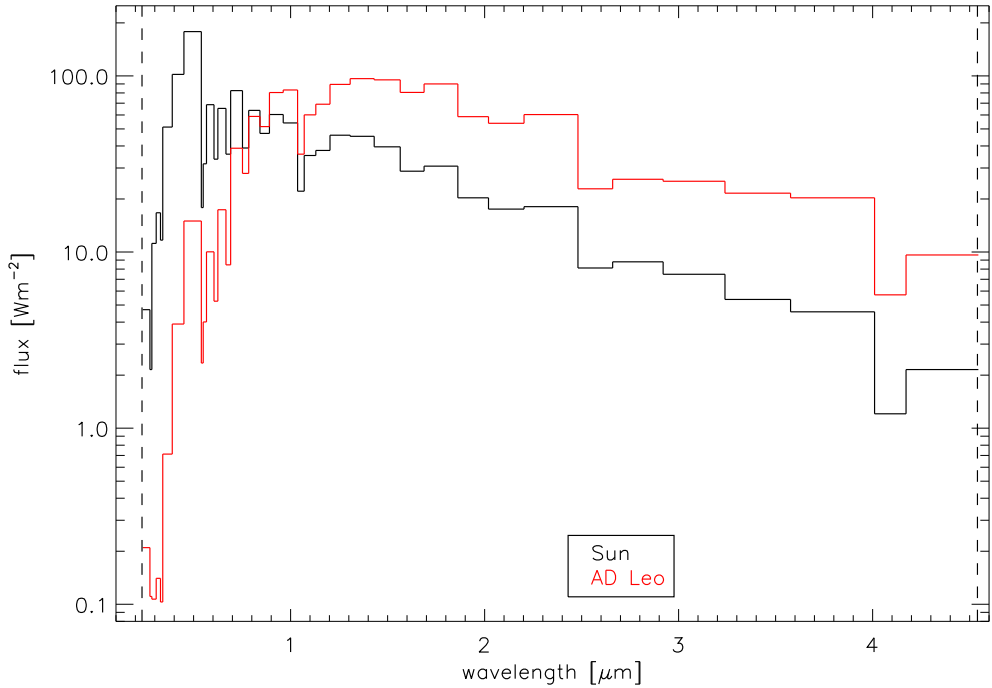}}\\
  \caption{Stellar spectra of the Sun and AD Leo as used in this work, normalized to equal modern-day Earth insolation. Vertical dashed lines indicate the spectral boundaries of the radiative transfer scheme (0.237-4.545\,$\mu$m, see Sect. \ref{radtra}).}
  \label{stellar}
\end{figure}

The planetary albedo $A_{p}$ is calculated as a Bond albedo in the model, i.e.

\begin{equation}
\label{bondalb}
A_{p}=\frac{F_{u}}{F_{\ast}}=\frac{\int^{\infty}_{0}A_{p,\lambda}\cdot F_{\ast,\lambda}d\lambda}{\int^{\infty}_{0} F_{\ast,\lambda}d\lambda}
\end{equation}
where $F_{u}$ is the spectrally integrated upwelling shortwave flux at the top of the atmosphere, $F_{\ast}$ the total stellar insolation, $A_{p,\lambda}$ the spectral albedo and $F_{\ast,\lambda}$ the spectral stellar insolation. As can be seen easily from this formulation, the planetary albedo of a given atmosphere is independent of orbital distance. $F_{\ast,\lambda}$ was approximated by stellar spectra for the Sun (as a G-star proxy) and AD Leo (representing an M-type star, with 0.41 solar radii, 0.4-0.5 solar masses and an effective temperature of $\approx$3,400 K, see e.g. \citealp{leggett1996}, \citealp{Seg2005}). Spectra were taken from \citet{kitzmann2010} and are shown in Fig. \ref{stellar}. The total stellar insolation is scaled to modern-day Earth insolation (i.e., $F_{\ast}$=1,366\,Wm$^{-2}$).

To investigate the influence of different atmospheric and surface properties on planetary albedo, we prescribed atmospheric composition, the surface pressure and the temperature profile for our calculations. Unless otherwise specified, the temperature profile was fixed at the modern-Earth profile $T_{\rm{m}}(z)$ ($z$ altitude), taken from \citet{grenfell2011}. The stellar zenith angle was fixed at 60$^{\circ}$ to simulate globally averaged conditions.

\subsection{Parameter variations}

\begin{figure}[h]
    \resizebox{\hsize}{!}{\includegraphics*{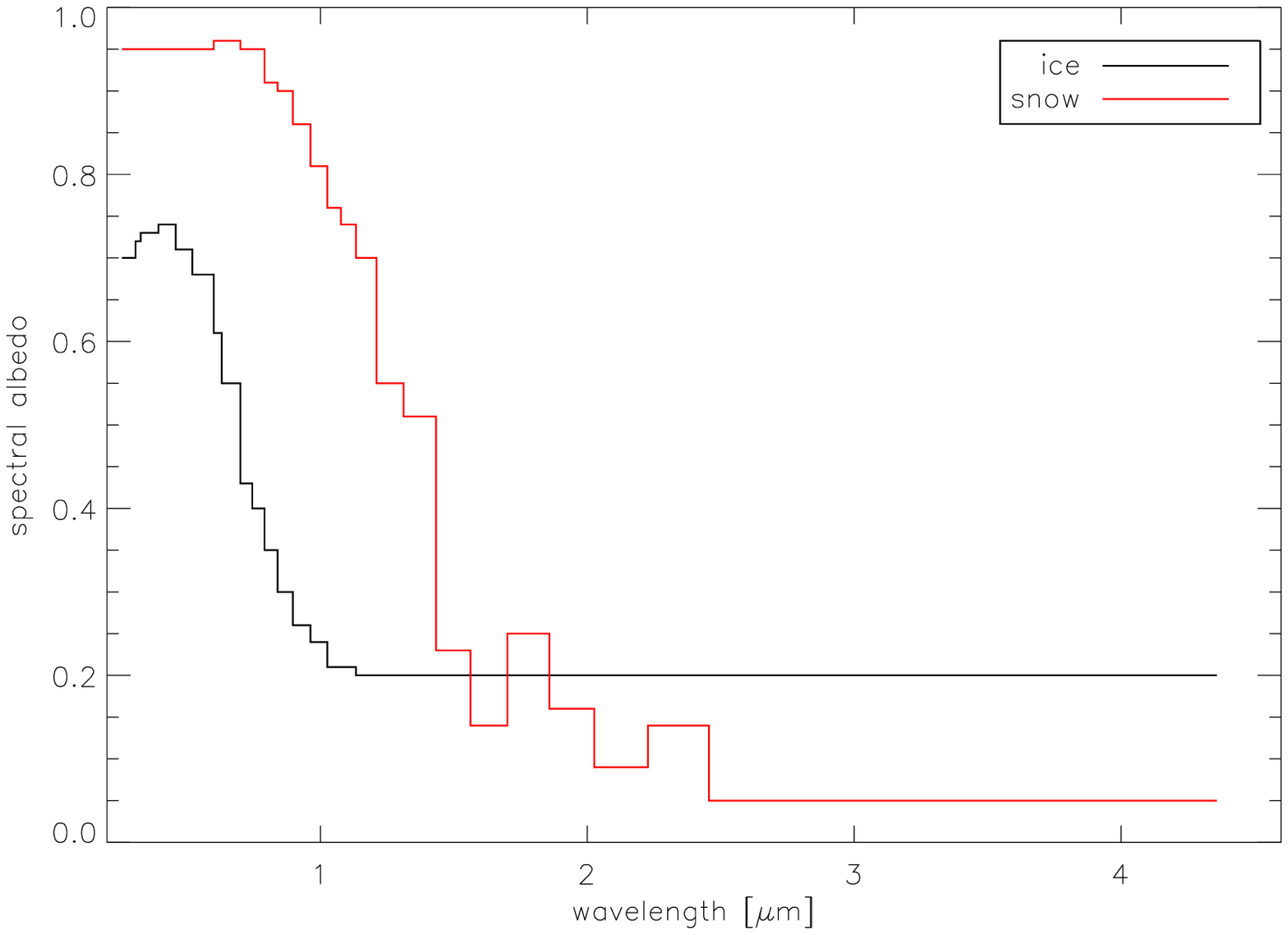}}\\
  \caption{Snow and ice spectral albedo. Taken from \citet{joshi2012}, their Fig. 1, data from \citet{brandt2005} and \citet{hudson2006}.}
  \label{icealb}
\end{figure}

We used two different types of surface albedo parameterizations. The first parametrization employed the wavelength-dependent surface albedos $A_S$ of ice and snow, as shown in Fig. \ref{icealb}. We took spectral albedo data presented in \citet{joshi2012}, which were originally obtained from measurements in Antarctica (\citealp{brandt2005}, \citealp{hudson2006}). Note that the albedo remains constant at wavelengths $\gtrsim$1.5\,$\mu$m (for ice, \citealp{brandt2005}) and $\gtrsim$2.4\,$\mu$m (for snow, \citealp{hudson2006}) since respective measurements were restricted to these spectral ranges. The second parametrization used a constant (wavelength-independent) surface albedo (values of 0.1, 0.2, 0.7 and 0.9, with 0.1 being close to present Earth's mean surface albedo, \citealp{rossow1999}).

The most important parameter for determining the location of the outer edge of the habitable zone is the CO$_{2}$ partial pressure. We performed calculations for 22 CO$_{2}$ partial pressures ranging between 0.1 and 30\,bar.

Since H$_{2}$O has very strong visible and near-IR absorption bands, even low concentrations could have an important impact on calculated albedos. To investigate this further, we performed simulations with the ice and snow surface albedos and atmospheres containing varying amounts of H$_2$O. H$_2$O concentrations $c_{\rm{H_2O}}$ as a function of altitude $z$ were set to follow the temperature-dependent saturation vapor pressure $p_{\rm{vap,H_2O}}(T)$, i.e. model atmospheres were assumed to be fully saturated:

\begin{equation}\label{waterprof}
  c_{\rm{H_2O}}(z)=\frac{p_{\rm{vap,H_2O}}(T(z))}{p(z)}
\end{equation}
Atmospheric temperatures $T(z)$ were set to the modern-Earth profile (i.e., $T(z)=T_{\rm{m}}(z)$, see Sect. \ref{procedure}) and to a colder profile with $T(z)=T_{\rm{m}}(z)-15$ (corresponding to a surface temperature of 273\,K). The colder temperature profile resulted in an approximate decrease in H$_2$O column amount of about a factor of 5-6.

CH$_{4}$ can build up in the atmosphere if strong surface fluxes exist, as postulated for, say, early Earth (e.g., \citealp{pavlov2003}). Planets around M stars favor a build-up of CH$_{4}$ because of the weak stellar UV radiation field and the related HOx chemistry (less OH means more CH$_{4}$, see e.g. \citealp{Seg2005} or \citealp{Grenf2007asbio}). CH$_{4}$ could have an important effect on the planetary albedo because of strong near-IR absorption bands. To investigate this, we performed calculations with CH$_{4}$ concentrations set to an isoprofile with concentrations of 10$^{-6}$, 10$^{-4}$ and 10$^{-3}$, respectively (again, with the ice and snow surface albedos).

O$_{3}$ can build up abiotically in CO$_{2}$-rich or strongly UV-irradiated atmospheres to quite large concentrations (e.g., \citealp{selsis2002}, \citealp{segura2007}, \citealp{domagal2010}) and could furthermore be the by-product of an oxygen-producing biosphere. It features strong absorption bands in the UV and visible part of the spectrum where Rayleigh scattering is strongest. Therefore, we performed sensitivity studies where we introduced an artificial ozone profile in the model atmospheres (1x and 10x modern Earth's concentrations following a mean Earth profile from \citealp{grenfell2011}, with the ice and snow surface albedos). We note that 10x modern Earth levels most likely represent upper limits for the O$_3$ profile. However since we aim at performing sensitivity studies only, we take this case as an end-member case.

All the simulations were done for an AD Leo and a Sun spectrum (see Fig. \ref{stellar}). Table \ref{planpar} summarizes the scenarios.

\begin{table*}
  \caption{Parameters for the model calculations performed (vmr volume mixing ratio, PAP present atmospheric profile, $w$: warm temperature profile $T_{\rm{m}}$, $c$: cold temperature profile $T_{\rm{m}}-15$, see Eq. \ref{waterprof}).}\label{planpar}
  \begin{tabular}{llcccc}
     \hline
   \hline
    Scenario & $A_S$ & $p_{\rm{CO_{2}}}$ [bar]& H$_2$O& CH$_{4}$ vmr & O$_{3}$ [PAP]\\
    \hline
    \hline
   1             &snow, ice (Fig. \ref{icealb})& 0.1, 0.2,..,1, 2,..,10,..,30 & 0  &0        &0\\
   2             &0.1, 0.2, 0.7, 0.9& 0.1, 0.2,..,1, 2,..,10,..,30 & 0  &0        &0\\
   3             &snow, ice (Fig. \ref{icealb})&0.1, 0.2,..,1, 2,..,10,..,30 & $w$, $c$ &0        &0\\
   4             &snow, ice (Fig. \ref{icealb})&0.1, 0.2,..,1, 2,..,10,..,30 & 0  & 10$^{-6}$,10$^{-4}$,10$^{-3}$       &0\\
   5            &snow, ice (Fig. \ref{icealb})& 0.1, 0.2,..,1, 2,..,10,..,30 & 0  &0        &1,10\\
   \hline
          \end{tabular}
\end{table*}

Note that we did not include N$_2$ in our model atmospheres, despite the fact that it is present in substantial amounts in all terrestrial atmospheres of the solar system (i.e., Venus, Earth, Mars, Titan). The reason is that N$_2$ does not feature absorption bands or lines in the visible and near-IR which are strong enough to affect the albedo and its Rayleigh scattering coefficient is nearly a factor of three smaller than for CO$_2$ \citep{vardavas1984}.

\label{para}

\section{Results}

\label{results}

\subsection{Atmospheric flux profiles}

\begin{figure}[h]
  \resizebox{\hsize}{!}{\includegraphics*{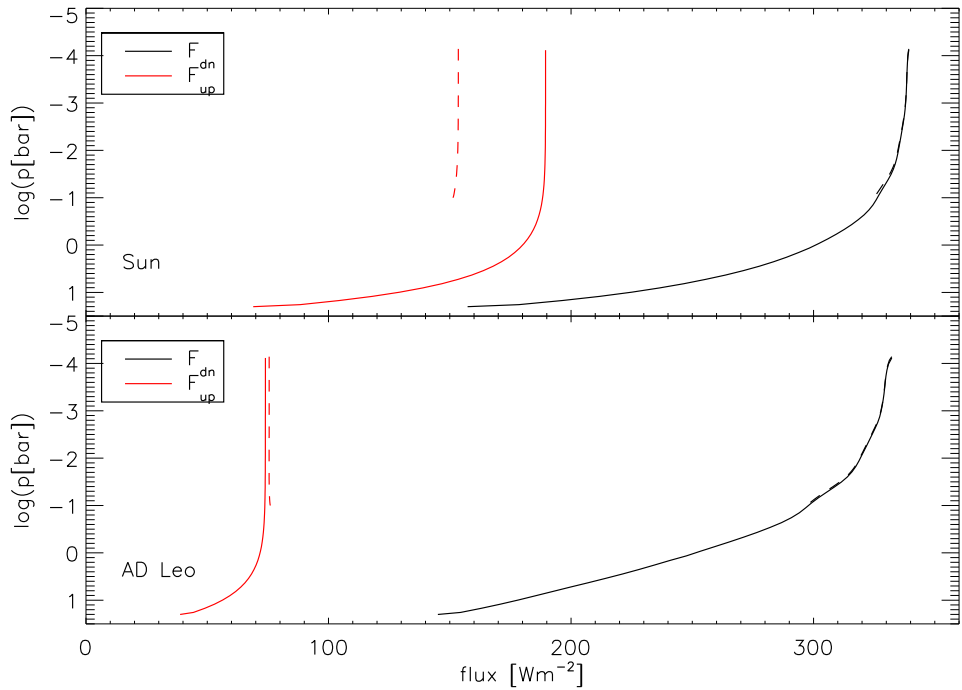}}\\
  \caption{Net upward (F$_{\rm{up}}$) and downward (F$_{\rm{dn}}$) fluxes in the visible/near-IR for ice surface albedo (scenario 1 from Table \ref{planpar}). Pure CO$_2$ atmospheres (20\,bar plain lines, 0.1\,bar dashed lines).}
  \label{netflux}
\end{figure}

Figure \ref{netflux} shows the downwards and upwards flux profiles in the visible/near-IR spectral domain as a function of pressure for the 0.1\,bar and the 20\,bar ice case of scenario 1 from Table \ref{planpar}. It is clearly seen that in the case of a 0.1\,bar CO$_2$ atmosphere, the influence of the atmosphere on these radiative fluxes, hence the influence on the planetary albedo, is rather small. The profiles all remain approximately constant throughout the entire atmosphere in this case. In contrast to that, in the 20\,bar scenario, the atmospheric influence is very strong.

For the planet around the Sun, the incoming stellar flux is roughly reduced by a factor of 2 (black line, 340\,Wm$^{-2}$ incoming flux at the top of the atmosphere, $\sim$ 160\,Wm$^{-2}$ reaching the surface). Most of this radiation is reflected back to space due to very efficient Rayleigh scattering (red line). The surface albedo contributes around 70\,Wm$^{-2}$ to the upwards flux at the top of the atmosphere, while the Rayleigh scattering component amounts to 120\,Wm$^{-2}$.

For the planet around the M-dwarf AD Leo, the influence of the atmosphere in the 20\,bar scenario is also very prominent. However, in this case it is primarily due to the absorption of near-IR radiation by CO$_2$. The surface and the Rayleigh scattering both contribute with about 50\,\% to the overall outgoing flux, but in absolute terms, the Rayleigh scattering component is reduced to about 35\,Wm$^{-2}$ (i.e., almost a factor of 4 compared to the Sun). The stellar flux reaching the surface is reduced from 340\,Wm$^{-2}$ to less than 150\,Wm$^{-2}$, which is only slightly less than for the planet around the Sun. Comparing the Rayleigh scattering component with the reduction in downwards flux, one finds that approximately 150\,Wm$^{-2}$ are absorbed in the atmosphere.

It is noteworthy that although the planetary atmosphere has a huge impact on the flux profiles, in the case of the planet around AD Leo the planetary albedo does not change that much. In the 0.1\,bar and the 20\,bar scenario, both top-of-atmosphere outgoing flux values are nearly identical.

\subsection{Planetary spectral albedo}

\label{planspecalb}

\begin{figure}[h]
   \resizebox{\hsize}{!}{ \includegraphics*{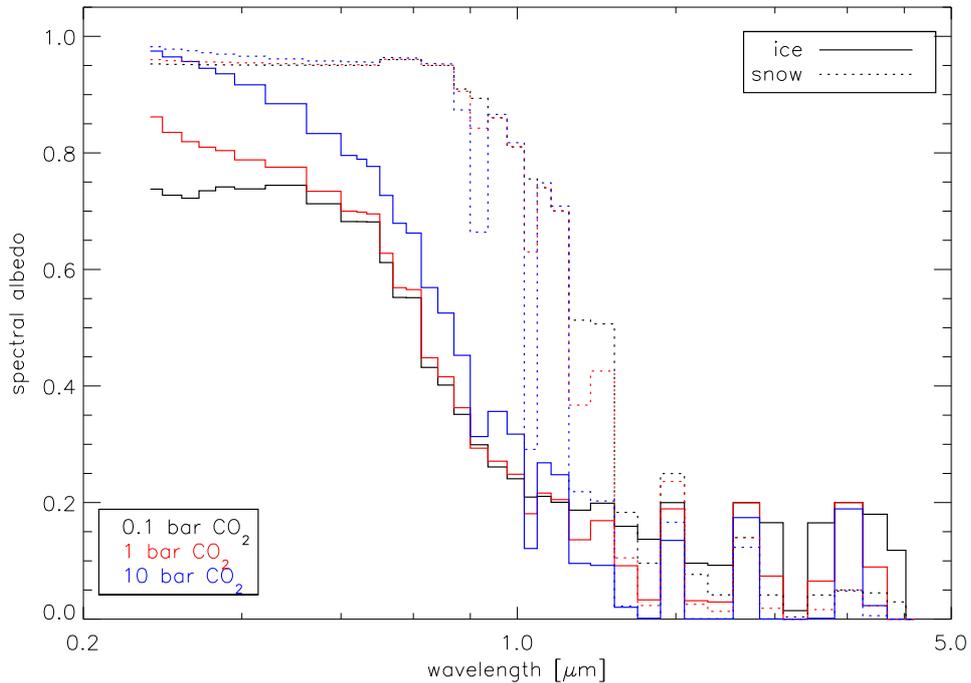}}\\
  \caption{Spectral albedos for different values of CO$_{2}$ pressure and ice and snow surface albedos. Pure CO$_2$ atmospheres.}
  \label{specalb}
\end{figure}

Figure \ref{specalb} shows the spectral albedos for three different values of the CO$_{2}$ pressure of scenario 1 from Table \ref{planpar}. It is clearly seen that an increase of CO$_{2}$ increases the spectral albedo in the visible (due to Rayleigh scattering) and decreases the spectral albedo in the near-IR (due to enhanced absorption). In general, the behavior of the spectral albedo follows the snow and ice spectral albedo in Fig. \ref{icealb}, i.e. a high spectral albedo in the visible and a rather low spectral albedo in the near-IR.

An important difference between ice and snow scenarios occurs near 1\,$\mu$m where the calculated albedos differ by a factor of about 3. This spectral region is important since a rather large part of stellar radiation is emitted there.

Note that for the visible part of the spectrum, where Rayleigh scattering is important, an increase of CO$_2$ from 0.1\,bar to 10\,bar does not significantly affect the spectral albedos for the snow surface case. In contrast, for the ice surface case, the effect is clearly distinguished. This is due to the very high visible albedo of snow ($\sim$0.9) in this spectral range, hence the surface reflection dominates the spectral albedo in the visible, independent of atmospheric conditions. The calculated values of the spectral albedos correspond well to values presented by \citet{shields2013}.

\subsection{Planetary  albedo}

\begin{figure}[h]
    \resizebox{\hsize}{!}{\includegraphics*{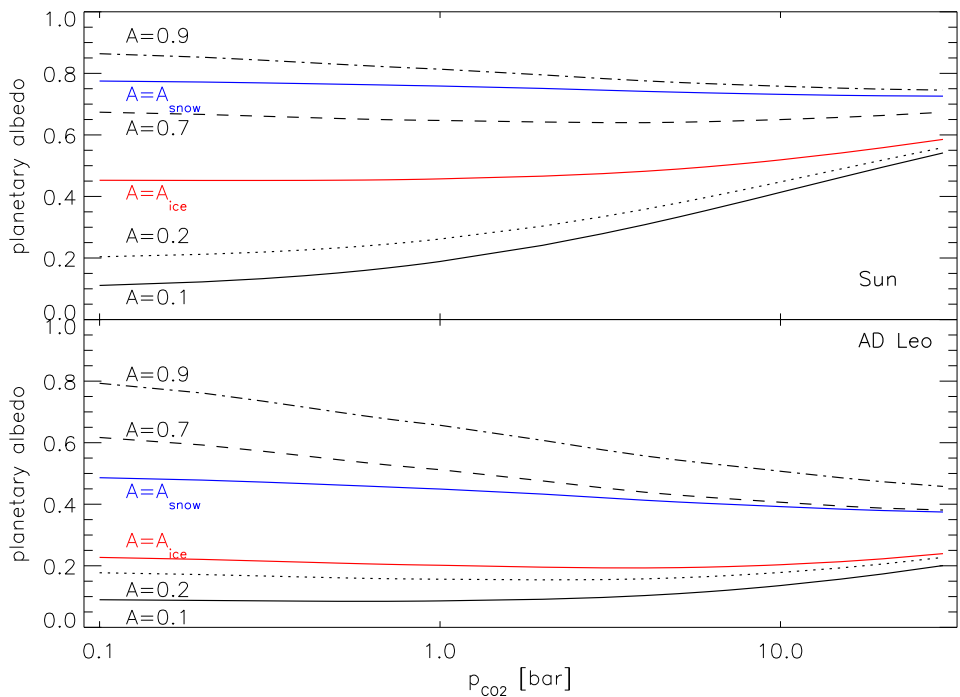}}\\
  \caption{Planetary albedo as a function of CO$_{2}$ pressure for scenarios 1 and 2 of Table \ref{planpar}. Pure CO$_2$ atmospheres.}
  \label{snow}
\end{figure}

Figure \ref{snow} shows the planetary Bond albedos of scenarios 1 and 2 from Table \ref{planpar}. The calculated values of planetary albedo with a very teneous CO$_{2}$ atmosphere (0.1\,bar) approximately confirm the values of \citet{joshi2012} which were done without taking into account atmospheric effects (their Figure 2).

It is clearly seen that the central star type has a large effect on the planetary albedo. At high CO$_{2}$ partial pressures, planetary albedos for the planet around the Sun are 2-3 times higher than for the planet orbiting around AD Leo. This is of course due to the different spectral distribution of stellar energy (see Fig. \ref{stellar}). In terms of energy-equivalent orbital distance, this implies that planets around M-stars could be 10-20\,\% farther away from their central star and still receive the same net stellar energy input into the atmosphere. This fact is the basis for the claim of \citet{joshi2012} that the HZ around M-stars is widened with respect to the HZ around G-stars.

As expected, planetary albedos are higher for snow surface albedo than for ice surface albedo. However, with increasing CO$_{2}$ partial pressure, the difference becomes noticeably smaller. For very dense, almost entirely optically thick atmospheres, planetary albedos will converge to a value independent of surface albedo.

Interestingly, the amount of CO$_{2}$ has a rather weak effect on the planetary albedo when using ice and snow albedos. This was already implied in Figs. \ref{netflux} and \ref{specalb}. The behavior of the atmospheric spectral albedo is similar to the spectral albedo of ice and snow (see Fig. \ref{icealb}). As before in Sect. \ref{planspecalb}, calculated values of Bond albedos correspond well to values in \citet{shields2013}.

However, for a wavelength-independent surface albedo values, the effect of CO$_{2}$ on planetary albedo is much stronger, especially for low (Sun) or high (AD Leo) values of surface albedo. This is explained as follows. Scattering alone can only increase the albedo. Thus, in the case of a Sun-like irradiation, for which scattering-only is a fair approximation, the planetary albedo increases with pressure. Therefore, for a given range of pressure, the increase is more significant for low surface albedos. Absorption alone can only decrease the albedo. Thus, for AD Leo-like irradiation, for which absorption-only is a fair approximation, the planetary albedo decreases with pressure. The effect is more significant if the surface albedo is high. From Fig. \ref{snow} it is clear that the scattering-alone and absorption-alone approximations are not perfect. For the Sun and high surface albedos, Fig. \ref{snow} shows a slight decrease of planetary albedo when surface pressure increases which is due to CO$_2$ absorption in the near IR. For AD Leo and low surface albedos, the planetary albedo increases with pressure due to the scattering that becomes more significant.

\subsubsection*{Effect of atmospheric composition}

\begin{figure}[h]
 \resizebox{\hsize}{!}{   \includegraphics*{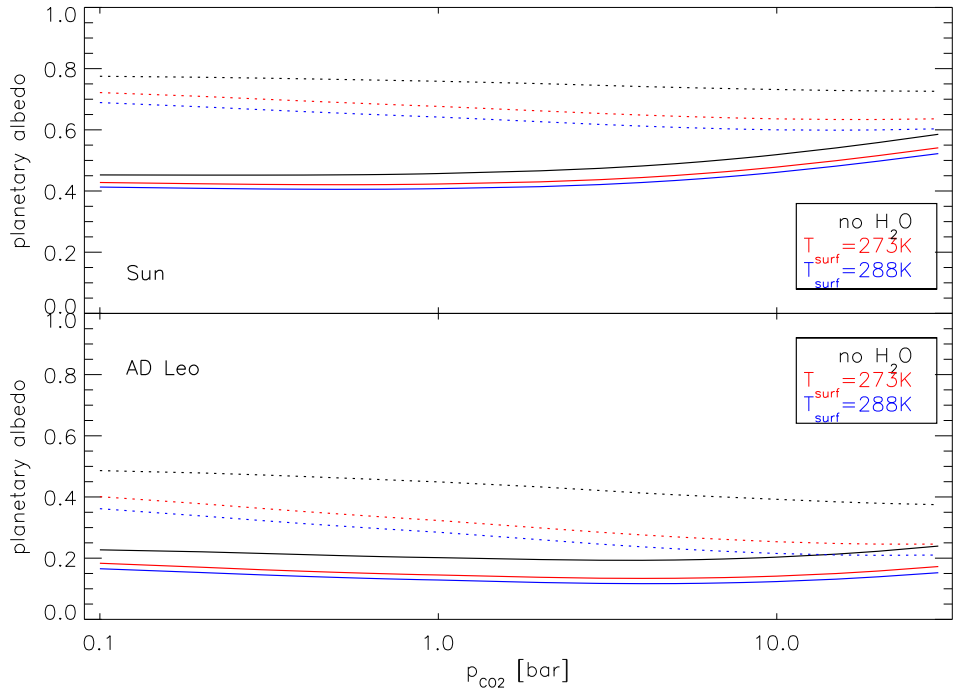}}\\
  \caption{Planetary albedo as a function of CO$_{2}$ pressure for snow and ice surface albedos. Influence of H$_{2}$O content. Plain lines for ice surface albedo, dotted lines for snow surface albedo.}
  \label{h2o}
\end{figure}

Fig. \ref{h2o} shows the effect of increasing H$_{2}$O concentrations on the calculated albedo. It is clearly seen that this decreases the planetary albedo. The effect is somewhat stronger for AD Leo than for the Sun. For example, for the ice surface albedo at 1\,bar of CO$_{2}$, when using the modern-day temperature profile and corresponding H$_2$O profile (see Eq. \ref{waterprof}), the planetary albedo decreases from 0.2 to 0.13 for AD Leo, but only from 0.45 to 0.40 for the Sun. This shows that even small amounts of water can enhance the effect of the stellar type on planetary albedo.

\begin{figure}[h]
    \resizebox{\hsize}{!}{\includegraphics*{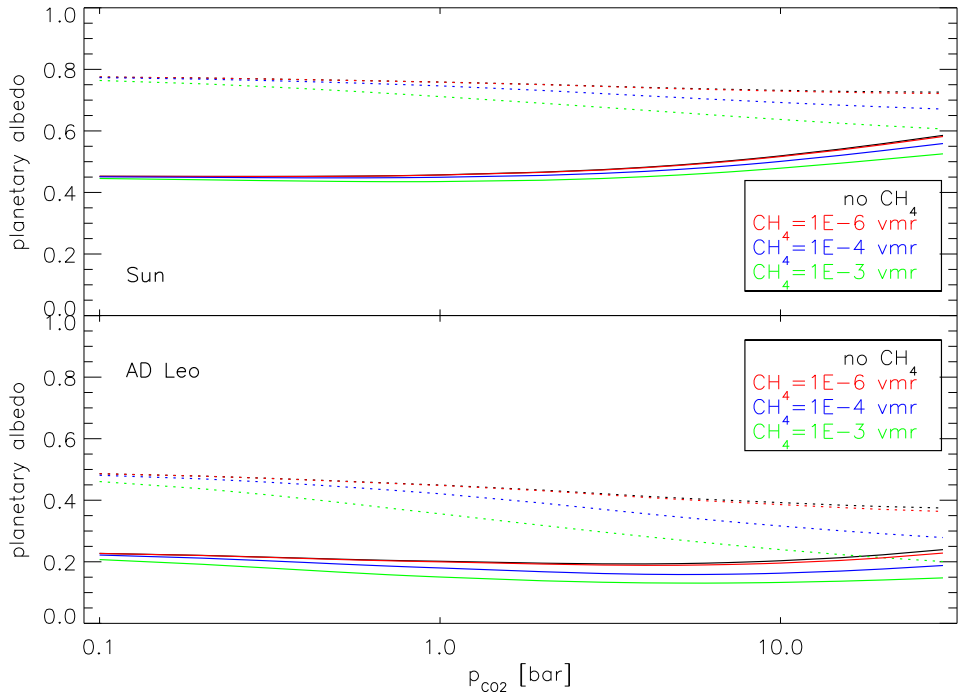}}\\
  \caption{Planetary albedo as a function of CO$_{2}$ pressure for snow and ice surface albedos. Influence of CH$_{4}$ content. Plain lines for ice surface albedo, dotted lines for snow surface albedo.}
  \label{meth}
\end{figure}

Fig. \ref{meth} shows the effect of increasing CH$_{4}$ concentration on the calculated albedo. Again, the effect is much stronger for AD Leo than for the Sun.

\begin{figure}[h]
 \resizebox{\hsize}{!}{   \includegraphics*{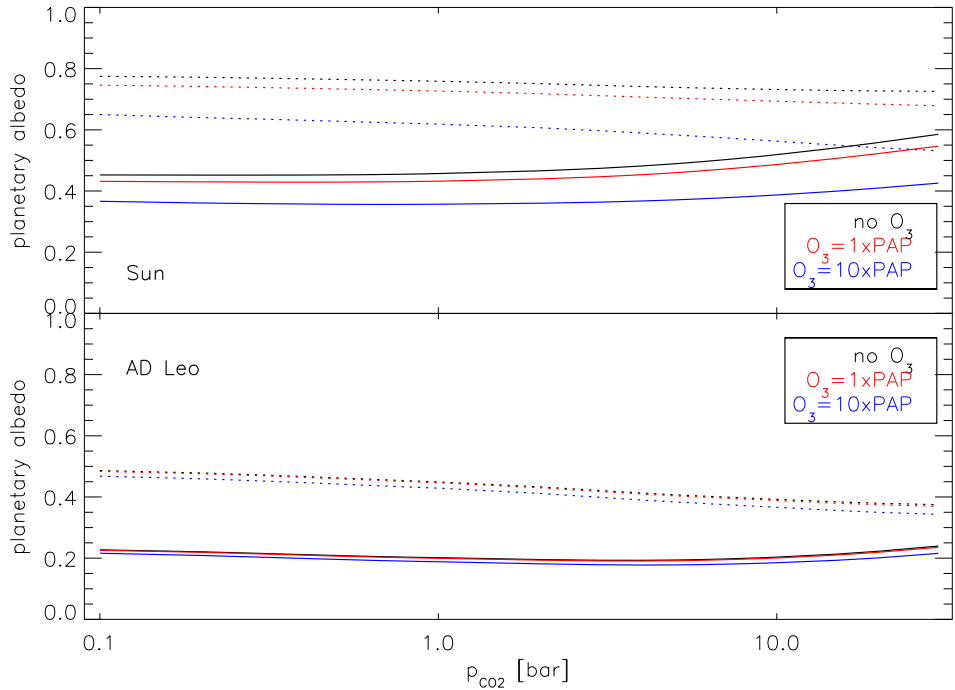}}\\
  \caption{Planetary albedo as a function of CO$_{2}$ pressure for snow and ice surface albedos. Influence of O$_{3}$ content. Plain lines for ice surface albedo, dotted lines for snow surface albedo.}
  \label{ozo}
\end{figure}

Fig. \ref{ozo} shows the effect of increasing O$_{3}$ concentration on the calculated albedo. It has a large impact for the Sun, but its effect is almost negligible for AD Leo, as expected from the spectral distribution of the stellar energy (see Fig. \ref{stellar}). O$_3$ most strongly absorbs at wavelengths below 0.5\,$\mu$m where AD Leo does not emit much radiation, compared to the Sun.

\section{Discussion of additional aspects}

\label{discuss}


\subsection{The effect of clouds}
 The main effect of clouds, such as in particular H$_2$O droplet and ice as well as CO$_2$ ice clouds, in the visible and NIR is the scattering of incident stellar radiation. Due to their high and continuous opacities they can easily dominate the planetary albedo in this wavelength region.

Additionally, clouds can also limit the impact of the scattering by  molecules and the surface on the planetary albedo. Shortwave radiation which is scattered upwards by gas molecules or the surface can in turn be scattered down again by the cloud. Thus, the amount of downwelling short-wave radiation in the lower atmosphere can be higher if a cloud is
present compared to the cloud-free cases presented in this study.

The exact impact on the incident stellar radiation is a direct function of the cloud's particle sizes. Particles with sizes near 0.1\,$\mu$m will have a Rayleigh scattering-like effect, i.e. they will contribute to the ice-albedo feedback in the same way as the scattering by gas molecules presented in this study. On the other hand, the optical properties of particles with sizes larger than a few $\mu$m are essentially flat in the visible and NIR (e.g., \citealp{kitzmann2010}). Compared to the cloud-free cases, this would lead to a constant offset in the planetary albedo which is almost independent from the incident stellar radiation (see e.g., \citealp{kitzmann2011reflect} for H$_2$O clouds and \citealp{kitzmann2013} for CO$_2$ clouds).

However, the presence of CO$_2$, H$_2$O, CH$_4$ and O$_3$ might inhibit cloud formation because of the warming of atmospheric layers by absorption of stellar radiation. Detailed atmospheric modeling is warranted to assess the possibility of cloud formation in the middle and upper atmosphere in future.

\subsection{Land-ocean distribution}

On Earth, the mean surface albedo is $A_S$=0.13 \citep{rossow1999}, with the oceans ($A_{\rm{ocean}}$=0.05) being less reflective than the continents. The Martian surface albedo is much higher than this ($A_{\rm{mars}}$=0.21, \citealp{kieffer1977}). Also, different types of land (deserts, bare rock, forests, etc.) have different spectral surface albedos. Land plants have emerged not earlier than about 600 million years ago on Earth (e.g., \citealp{kasthoward2006}), which demonstrates that profound changes of a planets surface albedo might happen during its evolution. Therefore, it is expected that the land-ocean ratio and the distribution of continents might have a large influence on the overall surface albedo, and by extension, on the sensitivity of the planet to the ice-albedo feedback.

Since this land-ocean ratio and the continental distribution might also change during the planetary evolution (e.g., \citealp{pesonen2012}), the strength of the ice-albedo effect might also change quite a lot over the course of the planetary evolution. An investigation with 2D or 3D atmosphere models would be necessary to assess the strength of these effects.

\subsection{Albedo measurements}

Surface and atmospheric properties of terrestrial exoplanets such as continents, oceans or ice, snow, and cloud coverage induce photometric variability in the visible-near IR light curves and affect measured secondary eclipse albedos. Recent studies have suggested that it might be possible to infer such properties from spectro-photometric observations (e.g., \citealp{oakley2009}, \citealp{fujii2010}, \citealp{livengood2011}, \citealp{cowan2011}). Albedo measurements might be indicative of snowball planets (e.g., \citealp{cowan2011}). Such observations are easiest for transiting planets and could be performed near secondary eclipse. For hot giant planets, this has already been done (e.g., \citealp{snellen2009}, \citealp{borucki2009}, \citealp{parviainen2013}).


Even if they were feasible for terrestrial, potentially habitable planets, it is questionable whether broadband photometric measurements of the reflected light could be useful to constrain the surface albedo and atmospheric characteristics. 
The contrast $C_R$ between star and planet in reflected light close to secondary eclipse depends on planetary radius $R_P$, planetary albedo $A_P$ and orbital distance $d$:

\begin{equation}\label{refl}
  C_R=A_P \cdot f_R= A_P \cdot \left(\frac{R_P}{d}\right)^2
\end{equation}
 For an Earth around the Sun ($d$=1\,AU), this translates into $C_R\sim A_P\cdot 1.8\times 10^{-9}$ (e.g., \citealp{kitzmann2011reflect}). If the planetary radius is known, the only uncertainty is on the deduced planetary albedo. Assuming a photometric precision of about 2$\times$10$^{-10}$ (i.e., 0.1$\times$$f_R$ in Eq. \ref{refl}, \citealp{oakley2009}), the 2\,$\sigma$ uncertainty for the obtained planetary albedo $A_p$ translates roughly to $\pm$0.2.


As shown above in Sect. \ref{results}, albedo differences for the various scenarios are generally of this order of magnitude. This implies that albedo characterization is very difficult for planets orbiting around M stars. Distinguishing between low surface albedos and completely ice-covered planets is unlikely, even if the high albedo due to snow coverage could be excluded reasonably well. Furthermore, it is very difficult to infer the atmospheric properties (i.e., surface pressure, amount of radiative trace gases). For planets orbiting around Sun-like stars, it may be possible to distinguish between ice-free and ice-covered planets at low CO$_2$ pressures. If a high albedo would be measured, however, the albedo solution is no longer unique, i.e. it might be an ice-covered planet at low or high atmospheric pressure or even an ice-free planet with a dense CO$_2$ atmosphere or clouds. 



Therefore, the characterization of the atmosphere and surface is not likely to be possible with secondary eclipse observations. To break degeneracies, additional observations are needed. Especially transmission spectroscopy during primary transit could be used to put constraints on atmospheric scale height and surface pressure (e.g., \citealp{vparis2011}, \citealp{benneke2012}), thus helping to interpret measured planetary albedos. These observations would, however, require large telescope facilities to reach sufficient S/N ratios (e.g., \citealp{hedelt2013})

\section{Summary and Conclusions}

We have presented detailed calculations of the planetary albedo of terrestrial planets as a function of stellar type and atmospheric composition. Key atmospheric properties such as CO$_2$ partial pressure and the abundance of radiative trace gases such as H$_2$O, CH$_4$, and O$_3$ were varied. Furthermore, we considered different types of surface albedo (wavelength-dependent ice or snow albedo, wavelength-independent constant albedo). Based on the planetary albedo, we investigated whether the ice-albedo feedback depends strongly on stellar type and atmospheric composition.

First, our calculations confirm the results of \citet{joshi2012}, i.e. that for ice- and snow-covered surfaces, the planetary albedos for planets orbiting M stars are much lower than for planets orbiting Sun-like stars (about a factor of two). This indeed leads to a strong reduction of the ice-albedo effect for planets around M stars.

Second, our results imply that for dense CO$_2$ atmospheres, the difference of planetary albedo between ice and ice-free ($A_S$=0.1) cases is strongly reduced (from 0.35 to 0.05 for planets orbiting the Sun). For planets around M stars, the ice-albedo feedback is almost entirely masked at high CO$_2$ pressures (albedo change less than 0.04).

Third, we have shown that the effect of atmospheric composition may be very important when calculating the strength of the ice-albedo effect. Even radiative trace species such as H$_2$O, CH$_4$, or O$_3$ could alter sensitively the planetary albedo. This in turn influences the radiative forcing associated with an ice or snow surface albedo, hence could strongly suppress the ice-albedo feedback. The presence of small amounts of H$_2$O and CH$_4$ effectively weakens the ice-albedo effect by several percent in planetary albedo for both stellar types considered. In addition, for planets around Sun-like stars, significant amounts of O$_3$ could also lead to a \textbf{very strong reduction} of the ice-albedo feedback at high CO$_2$ pressures.

\label{summary}

\section*{Acknowledgements}

P. v. Paris and F. Selsis acknowledge support from the European Research
Council (Starting Grant 209622: E$_3$ARTHs). This research has been partly supported by the Helmholtz Association through the research alliance "Planetary Evolution and Life". Valuable discussions with P. Hedelt and J.L. Grenfell are gratefully acknowledged. Helpful, constructive comments by D. Abbot and an anonymous reviewer are also acknowledged.

\bibliographystyle{natbib}
\bibliography{literatur_joshi}

\begin{thebibliography}{49}
\providecommand{\natexlab}[1]{#1}
\providecommand{\url}[1]{\texttt{#1}}
\expandafter\ifx\csname urlstyle\endcsname\relax
  \providecommand{\doi}[1]{doi: #1}\else
  \providecommand{\doi}{doi: \begingroup \urlstyle{rm}\Url}\fi

\bibitem[{Abbot} et~al.(2012){Abbot}, {Cowan}, and {Ciesla}]{abbot2012}
D.~S. {Abbot}, N.~B. {Cowan}, and F.~J. {Ciesla}.
\newblock {Indication of Insensitivity of Planetary Weathering Behavior and
  Habitable Zone to Surface Land Fraction}.
\newblock \emph{\apj}, 756:\penalty0 178, September 2012.
\newblock \doi{10.1088/0004-637X/756/2/178}.

\bibitem[{Allen}(1973)]{allen1973}
C.W. {Allen}.
\newblock \emph{Astrophysical Quantities}.
\newblock The Athlone Press, University of London, 1973.

\bibitem[{Benneke} and {Seager}(2012)]{benneke2012}
B.~{Benneke} and S.~{Seager}.
\newblock {Atmospheric Retrieval for Super-Earths: Uniquely Constraining the
  Atmospheric Composition with Transmission Spectroscopy}.
\newblock \emph{\apj}, 753:\penalty0 100, July 2012.
\newblock \doi{10.1088/0004-637X/753/2/100}.

\bibitem[{Borucki} et~al.(2009){Borucki}, {Koch}, {Jenkins}, {Sasselov},
  {Gilliland}, {Batalha}, {Latham}, {Caldwell}, {Basri}, {Brown},
  {Christensen-Dalsgaard}, {Cochran}, {DeVore}, {Dunham}, {Dupree}, {Gautier},
  {Geary}, {Gould}, {Howell}, {Kjeldsen}, {Lissauer}, {Marcy}, {Meibom},
  {Morrison}, and {Tarter}]{borucki2009}
W.~J. {Borucki}, D.~{Koch}, J.~{Jenkins}, D.~{Sasselov}, R.~{Gilliland},
  N.~{Batalha}, D.~W. {Latham}, D.~{Caldwell}, G.~{Basri}, T.~{Brown},
  J.~{Christensen-Dalsgaard}, W.~D. {Cochran}, E.~{DeVore}, E.~{Dunham}, A.~K.
  {Dupree}, T.~{Gautier}, J.~{Geary}, A.~{Gould}, S.~{Howell}, H.~{Kjeldsen},
  J.~{Lissauer}, G.~{Marcy}, S.~{Meibom}, D.~{Morrison}, and J.~{Tarter}.
\newblock {Kepler's Optical Phase Curve of the Exoplanet HAT-P-7b}.
\newblock \emph{Science}, 325:\penalty0 709, August 2009.
\newblock \doi{10.1126/science.1178312}.

\bibitem[{Brandt} et~al.(2005){Brandt}, {Warren}, {Worby}, and
  {Grenfell}]{brandt2005}
R.~E. {Brandt}, S.~G. {Warren}, A.~P. {Worby}, and T.~C. {Grenfell}.
\newblock {Surface Albedo of the Antarctic Sea Ice Zone.}
\newblock \emph{Journal of Climate}, 18:\penalty0 3606--3622, September 2005.
\newblock \doi{10.1175/JCLI3489.1}.

\bibitem[{Cowan} et~al.(2011){Cowan}, {Robinson}, {Livengood}, {Deming},
  {Agol}, {A'Hearn}, {Charbonneau}, {Lisse}, {Meadows}, {Seager}, {Shields},
  and {Wellnitz}]{cowan2011}
N.~B. {Cowan}, T.~{Robinson}, T.~A. {Livengood}, D.~{Deming}, E.~{Agol}, M.~F.
  {A'Hearn}, D.~{Charbonneau}, C.~M. {Lisse}, V.~S. {Meadows}, S.~{Seager},
  A.~L. {Shields}, and D.~D. {Wellnitz}.
\newblock {Rotational Variability of Earth's Polar Regions: Implications for
  Detecting Snowball Planets}.
\newblock \emph{\apj}, 731:\penalty0 76, April 2011.
\newblock \doi{10.1088/0004-637X/731/1/76}.

\bibitem[{Domagal-Goldman} and {Meadows}(2010)]{domagal2010}
S.~{Domagal-Goldman} and V.~{Meadows}.
\newblock Abiotic buildup of ozone.
\newblock \emph{ASP Conference Series}, 430:\penalty0 152, 2010.

\bibitem[{Fujii} et~al.(2010){Fujii}, {Kawahara}, {Suto}, {Taruya}, {Fukuda},
  {Nakajima}, and {Turner}]{fujii2010}
Y.~{Fujii}, H.~{Kawahara}, Y.~{Suto}, A.~{Taruya}, S.~{Fukuda}, T.~{Nakajima},
  and E.~L. {Turner}.
\newblock {Colors of a Second Earth: Estimating the Fractional Areas of Ocean,
  Land, and Vegetation of Earth-like Exoplanets}.
\newblock \emph{\apj}, 715:\penalty0 866--880, June 2010.
\newblock \doi{10.1088/0004-637X/715/2/866}.

\bibitem[{Goldblatt} et~al.(2006){Goldblatt}, {Lenton}, and
  {Watson}]{goldblatt2006}
C.~{Goldblatt}, T.~M. {Lenton}, and A.~J. {Watson}.
\newblock {Bistability of atmospheric oxygen and the Great Oxidation}.
\newblock \emph{\nat}, 443:\penalty0 683--686, October 2006.
\newblock \doi{10.1038/nature05169}.

\bibitem[{Grenfell} et~al.(2007){Grenfell}, {Grie{\ss}meier}, {Patzer},
  {Rauer}, {Segura}, {Stadelmann}, {Stracke}, {Titz}, and {Von
  Paris}]{Grenf2007asbio}
J.~L. {Grenfell}, J.-M. {Grie{\ss}meier}, B.~{Patzer}, H.~{Rauer}, A.~{Segura},
  A.~{Stadelmann}, B.~{Stracke}, R.~{Titz}, and P.~{Von Paris}.
\newblock {Biomarker Response to Galactic Cosmic Ray-Induced NO$_x$ And The
  Methane Greenhouse Effect in The Atmosphere of An Earth-Like Planet Orbiting
  An M Dwarf Star}.
\newblock \emph{\astrb}, 7:\penalty0 208--221, February 2007.
\newblock \doi{10.1089/ast.2006.0129}.

\bibitem[{Grenfell} et~al.(2011){Grenfell}, {Gebauer}, {von Paris}, {Godolt},
  {Hedelt}, {Patzer}, {Stracke}, and {Rauer}]{grenfell2011}
J.~L. {Grenfell}, S.~{Gebauer}, P.~{von Paris}, M.~{Godolt}, P.~{Hedelt},
  A.~B.~C. {Patzer}, B.~{Stracke}, and H.~{Rauer}.
\newblock {Sensitivity of biomarkers to changes in chemical emissions in the
  Earth's Proterozoic atmosphere}.
\newblock \emph{\icarus}, 211:\penalty0 81--88, January 2011.
\newblock \doi{10.1016/j.icarus.2010.09.015}.

\bibitem[{Hedelt} et~al.(2013){Hedelt}, {von Paris}, {Godolt}, {Gebauer},
  {Grenfell}, {Rauer}, {Schreier}, {Selsis}, and {Trautmann}]{hedelt2013}
P.~{Hedelt}, P.~{von Paris}, M.~{Godolt}, S.~{Gebauer}, J.~L. {Grenfell},
  H.~{Rauer}, F.~{Schreier}, F.~{Selsis}, and T.~{Trautmann}.
\newblock {Spectral features of Earth-like planets and their detectability at
  different orbital distances around F, G, and K-type stars}.
\newblock \emph{\aap}, 553:\penalty0 A9, May 2013.
\newblock \doi{10.1051/0004-6361/201117723}.

\bibitem[{Hoffman} et~al.(1998){Hoffman}, {Kaufman}, {Halverson}, and
  {Schrag}]{hoffman1998}
P.~F. {Hoffman}, A.~J. {Kaufman}, G.~P. {Halverson}, and D.~P. {Schrag}.
\newblock {A Neoproterozoic Snowball Earth}.
\newblock \emph{Science}, 281:\penalty0 1342, August 1998.
\newblock \doi{10.1126/science.281.5381.1342}.

\bibitem[{Hudson} et~al.(2006){Hudson}, {Warren}, {Brandt}, {Grenfell}, and
  {Six}]{hudson2006}
S.~R. {Hudson}, S.~G. {Warren}, R.~E. {Brandt}, T.~C. {Grenfell}, and D.~{Six}.
\newblock {Spectral bidirectional reflectance of Antarctic snow: Measurements
  and parameterization}.
\newblock \emph{Journal of Geophysical Research (Atmospheres)}, 111:\penalty0
  D18106, September 2006.
\newblock \doi{10.1029/2006JD007290}.

\bibitem[{Hyde} et~al.(2000){Hyde}, {Crowley}, {Baum}, and {Peltier}]{hyde2000}
W.~T. {Hyde}, T.~J. {Crowley}, S.~K. {Baum}, and W.~R. {Peltier}.
\newblock {Neoproterozoic `snowball Earth' simulations with a coupled
  climate/ice-sheet model}.
\newblock \emph{\nat}, 405:\penalty0 425--429, May 2000.

\bibitem[{Joshi} and {Haberle}(2012)]{joshi2012}
M.~M. {Joshi} and R.~M. {Haberle}.
\newblock {Suppression of the Water Ice and Snow Albedo Feedback on Planets
  Orbiting Red Dwarf Stars and the Subsequent Widening of the Habitable Zone}.
\newblock \emph{Astrobiology}, 12:\penalty0 3--8, January 2012.
\newblock \doi{10.1089/ast.2011.0668}.

\bibitem[{Karkoschka}(1994)]{karkoschka1994}
E.~{Karkoschka}.
\newblock {Spectrophotometry of the jovian planets and Titan at 300- to 1000-nm
  wavelength: The methane spectrum}.
\newblock \emph{Icarus}, 111:\penalty0 174--192, September 1994.
\newblock \doi{10.1006/icar.1994.1139}.

\bibitem[{Kasting} and {Howard}(2006)]{kasthoward2006}
J.~F. {Kasting} and M.~T. {Howard}.
\newblock {Atmospheric composition and climate on the early Earth}.
\newblock \emph{Phil. Trans. R. Soc. B}, 361:\penalty0 1733--1742, 2006.

\bibitem[{Kasting} and {Ono}(2006)]{kastono2006}
J.~F. {Kasting} and S.~{Ono}.
\newblock {Palaeoclimates: the first two billion years}.
\newblock \emph{Phil. Trans. R. Soc. B}, 361:\penalty0 917--929, 2006.

\bibitem[{Kasting} et~al.(1993){Kasting}, {Whitmire}, and
  {Reynolds}]{kasting1993}
J.~F. {Kasting}, D.~P. {Whitmire}, and R.~T. {Reynolds}.
\newblock {Habitable Zones around Main Sequence Stars}.
\newblock \emph{\icarus}, 101:\penalty0 108--128, January 1993.
\newblock \doi{10.1006/icar.1993.1010}.

\bibitem[{Kieffer} et~al.(1977){Kieffer}, {Martin}, {Peterfreund}, {Jakosky},
  {Miner}, and {Palluconi}]{kieffer1977}
H.~H. {Kieffer}, T.~Z. {Martin}, A.~R. {Peterfreund}, B.~M. {Jakosky}, E.~D.
  {Miner}, and F.~D. {Palluconi}.
\newblock {Thermal and albedo mapping of Mars during the Viking primary
  mission}.
\newblock \emph{\jgr}, 82:\penalty0 4249--4291, September 1977.
\newblock \doi{10.1029/JS082i028p04249}.

\bibitem[{Kirschvink}(1992)]{kirschvink1992}
J.~{Kirschvink}.
\newblock {Late Proterozoic low-latitude global glaciation: The snowball Earth
  }.
\newblock In \emph{The Proterozoic Biosphere: A Multidisciplinary Study}, pages
  51--52, 1992.

\bibitem[{Kitzmann} et~al.(2010){Kitzmann}, {Patzer}, {von Paris}, {Godolt},
  {Stracke}, {Gebauer}, {Grenfell}, and {Rauer}]{kitzmann2010}
D.~{Kitzmann}, A.~B.~C. {Patzer}, P.~{von Paris}, M.~{Godolt}, B.~{Stracke},
  S.~{Gebauer}, J.~L. {Grenfell}, and H.~{Rauer}.
\newblock {Clouds in the atmospheres of extrasolar planets. I. Climatic effects
  of multi-layered clouds for Earth-like planets and implications for habitable
  zones}.
\newblock \emph{\aa}, 511:\penalty0 A66, March 2010.

\bibitem[{Kitzmann} et~al.(2011){Kitzmann}, {Patzer}, {von Paris}, {Godolt},
  and {Rauer}]{kitzmann2011reflect}
D.~{Kitzmann}, A.~B.~C. {Patzer}, P.~{von Paris}, M.~{Godolt}, and H.~{Rauer}.
\newblock {Clouds in the atmospheres of extrasolar planets. III. Impact of low
  and high-level clouds on the reflection spectra of Earth-like planets}.
\newblock \emph{\aa}, 534:\penalty0 A63, October 2011.
\newblock \doi{10.1051/0004-6361/201117375}.

\bibitem[{Kitzmann} et~al.(2013){Kitzmann}, {Patzer}, and
  {Rauer}]{kitzmann2013}
D.~{Kitzmann}, A.~B.~C. {Patzer}, and H.~{Rauer}.
\newblock {Clouds in the atmospheres of extrasolar planets. IV. On the
  scattering greenhouse effect of CO2 ice particles: Numerical radiative
  transfer studies}.
\newblock \emph{accepted in \aa}, June 2013.

\bibitem[{Leggett} et~al.(1996){Leggett}, {Allard}, {Berriman}, {Dahn}, and
  {Hauschildt}]{leggett1996}
S.~K. {Leggett}, F.~{Allard}, G.~{Berriman}, C.~C. {Dahn}, and P.~H.
  {Hauschildt}.
\newblock {Infrared Spectra of Low-Mass Stars: Toward a Temperature Scale for
  Red Dwarfs}.
\newblock \emph{\apjs}, 104:\penalty0 117--143, May 1996.
\newblock \doi{10.1086/192295}.

\bibitem[{Livengood} et~al.(2011){Livengood}, {Deming}, {A'Hearn},
  {Charbonneau}, {Hewagama}, {Lisse}, {McFadden}, {Meadows}, {Robinson},
  {Seager}, and {Wellnitz}]{livengood2011}
T.~A. {Livengood}, L.~D. {Deming}, M.~F. {A'Hearn}, D.~{Charbonneau},
  T.~{Hewagama}, C.~M. {Lisse}, L.~A. {McFadden}, V.~S. {Meadows}, T.~D.
  {Robinson}, S.~{Seager}, and D.~D. {Wellnitz}.
\newblock {Properties of an Earth-Like Planet Orbiting a Sun-Like Star: Earth
  Observed by the EPOXI Mission}.
\newblock \emph{Astrobiology}, 11:\penalty0 907--930, November 2011.
\newblock \doi{10.1089/ast.2011.0614}.

\bibitem[{McKay} et~al.(1989){McKay}, {Pollack}, and {Courtin}]{mckay1989titan}
C.~P. {McKay}, J.~B. {Pollack}, and R.~{Courtin}.
\newblock {The thermal structure of Titan's atmosphere}.
\newblock \emph{\icarus}, 80:\penalty0 23--53, July 1989.
\newblock \doi{10.1016/0019-1035(89)90160-7}.

\bibitem[{Mischna} et~al.(2000){Mischna}, {Kasting}, {Pavlov}, and
  {Freedman}]{mischna2000}
M.~A. {Mischna}, J.~F. {Kasting}, A.~{Pavlov}, and R.~{Freedman}.
\newblock {Influence of carbon dioxide clouds on early martian climate}.
\newblock \emph{\icarus}, 145:\penalty0 546--554, June 2000.
\newblock \doi{10.1006/icar.2000.6380}.

\bibitem[{Oakley} and {Cash}(2009)]{oakley2009}
P.~H.~H. {Oakley} and W.~{Cash}.
\newblock {Construction of an Earth Model: Analysis of Exoplanet Light Curves
  and Mapping the Next Earth with the New Worlds Observer}.
\newblock \emph{\apj}, 700:\penalty0 1428--1439, August 2009.
\newblock \doi{10.1088/0004-637X/700/2/1428}.

\bibitem[{Parviainen} et~al.(2013){Parviainen}, {Deeg}, and
  {Belmonte}]{parviainen2013}
H.~{Parviainen}, H.~J. {Deeg}, and J.~A. {Belmonte}.
\newblock {Secondary eclipses in the CoRoT light curves. A homogeneous search
  based on Bayesian model selection}.
\newblock \emph{\aap}, 550:\penalty0 A67, February 2013.
\newblock \doi{10.1051/0004-6361/201220081}.

\bibitem[{Pavlov} et~al.(2000){Pavlov}, {Kasting}, {Brown}, {Rages}, and
  {Freedman}]{pavlov2000}
A.~A. {Pavlov}, J.~F. {Kasting}, L.~L. {Brown}, K.~A. {Rages}, and
  R.~{Freedman}.
\newblock {Greenhouse warming by CH$_4$ in the atmosphere of early Earth}.
\newblock \emph{\jgr}, 105:\penalty0 11981--11990, May 2000.
\newblock \doi{10.1029/1999JE001134}.

\bibitem[{Pavlov} et~al.(2003){Pavlov}, {Hurtgen}, {Kasting}, and
  {Arthur}]{pavlov2003}
A.~A. {Pavlov}, M.~T. {Hurtgen}, J.~F. {Kasting}, and M.~A. {Arthur}.
\newblock {Methane-rich Proterozoic atmosphere?}
\newblock \emph{Geology}, 31:\penalty0 87--90, January 2003.
\newblock \doi{10.1130}.

\bibitem[{Pesonen} et~al.(2012){Pesonen}, {Mertanen}, and
  {Veikkolainen}]{pesonen2012}
L.~{Pesonen}, S.~{Mertanen}, and T.~{Veikkolainen}.
\newblock {Paleo-Mesoproterozoic Supercontinents – A Paleomagnetic View}.
\newblock \emph{Geophysica}, 48:\penalty0 5--47, January 2012.

\bibitem[{Rossow} and {Schiffer}(1999)]{rossow1999}
W.~B. {Rossow} and R.~A. {Schiffer}.
\newblock {Advances in Understanding Clouds from ISCCP.}
\newblock \emph{Bull. Americ. Meteor. Soc.}, 80:\penalty0 2261--2288, November
  1999.
\newblock \doi{10.1175/1520-0477(1999)080<2261:AIUCFI>2.0.CO;2}.

\bibitem[{Segura} et~al.(2005){Segura}, {Kasting}, {Meadows}, {Cohen}, {Scalo},
  {Crisp}, {Butler}, and {Tinetti}]{Seg2005}
A.~{Segura}, J.~F. {Kasting}, V.~{Meadows}, M.~{Cohen}, J.~{Scalo}, D.~{Crisp},
  R.~A.~H. {Butler}, and G.~{Tinetti}.
\newblock {Biosignatures from Earth-Like Planets Around M Dwarfs}.
\newblock \emph{\astrb}, 5:\penalty0 706--725, December 2005.
\newblock \doi{10.1089/ast.2005.5.706}.

\bibitem[{Segura} et~al.(2007){Segura}, {Meadows}, {Kasting}, {Crisp}, and
  {Cohen}]{segura2007}
A.~{Segura}, V.~S. {Meadows}, J.~F. {Kasting}, D.~{Crisp}, and M.~{Cohen}.
\newblock {Abiotic formation of O$_2$ and O$_3$ in high-CO$_2$ terrestrial
  atmospheres}.
\newblock \emph{\aa}, 472:\penalty0 665--679, 2007.

\bibitem[{Selsis} et~al.(2002){Selsis}, {Despois}, and {Parisot}]{selsis2002}
F.~{Selsis}, D.~{Despois}, and J.-P. {Parisot}.
\newblock {Signature of life on exoplanets: Can Darwin produce false positive
  detections?}
\newblock \emph{\aa}, 388:\penalty0 985--1003, June 2002.
\newblock \doi{10.1051/0004-6361:20020527}.

\bibitem[{Selsis} et~al.(2007){Selsis}, {Kasting}, {Levrard}, {Paillet},
  {Ribas}, and {Delfosse}]{selsis2007gliese}
F.~{Selsis}, J.~F. {Kasting}, B.~{Levrard}, J.~{Paillet}, I.~{Ribas}, and
  X.~{Delfosse}.
\newblock {Habitable planets around the star Gliese 581?}
\newblock \emph{\aa}, 476:\penalty0 1373--1387, December 2007.
\newblock \doi{10.1051/0004-6361:20078091}.

\bibitem[{Shields} et~al.(2013){Shields}, {Meadows}, {Bitz}, {Pierrehumbert},
  {Joshi}, and {Robinson}]{shields2013}
A.~L. {Shields}, V.~S. {Meadows}, C.~M. {Bitz}, R.~T. {Pierrehumbert}, M.~M.
  {Joshi}, and T.~D. {Robinson}.
\newblock {The Effect of Host Star Spectral Energy Distribution and Ice-Albedo
  Feedback on the Climate of Extrasolar Planets}.
\newblock \emph{accepted in \astrb}, May 2013.

\bibitem[{Snellen} et~al.(2009){Snellen}, {de Mooij}, and
  {Albrecht}]{snellen2009}
I.~A.~G. {Snellen}, E.~J.~W. {de Mooij}, and S.~{Albrecht}.
\newblock {The changing phases of extrasolar planet CoRoT-1b}.
\newblock \emph{\nat}, 459:\penalty0 543--545, May 2009.
\newblock \doi{10.1038/nature08045}.

\bibitem[{Tarasov} and {Peltier}(1997)]{tarasov1997}
L.~{Tarasov} and W.~R. {Peltier}.
\newblock {Terminating the 100 kyr ice age cycle}.
\newblock \emph{\jgr}, 102:\penalty0 21665--21694, 1997.
\newblock \doi{10.1029/97JD01766}.

\bibitem[{Toon} et~al.(1989){Toon}, {McKay}, {Ackerman}, and
  {Santhanam}]{toon1989}
O.~B. {Toon}, C.~P. {McKay}, T.~P. {Ackerman}, and K.~{Santhanam}.
\newblock {Rapid calculation of radiative heating rates and photodissociation
  rates in inhomogeneous multiple scattering atmospheres}.
\newblock \emph{\jgr}, 94:\penalty0 16287--16301, 1989.

\bibitem[{Vardavas} and {Carver}(1984)]{vardavas1984}
I.~M. {Vardavas} and J.~H. {Carver}.
\newblock {Solar and terrestrial parameterizations for radiative-convective
  models}.
\newblock \emph{\planss}, 32:\penalty0 1307--1325, October 1984.
\newblock \doi{10.1016/0032-0633(84)90074-6}.

\bibitem[{von Paris} et~al.(2010){von Paris}, {Gebauer}, {Godolt}, {Grenfell},
  {Hedelt}, {Kitzmann}, {Patzer}, {Rauer}, and {Stracke}]{vparis2010gliese}
P.~{von Paris}, S.~{Gebauer}, M.~{Godolt}, J.~L. {Grenfell}, P.~{Hedelt},
  D.~{Kitzmann}, A.~B.~C. {Patzer}, H.~{Rauer}, and B.~{Stracke}.
\newblock {The extrasolar planet GL 581 d: A potentially habitable planet?}
\newblock \emph{\aa}, 522:\penalty0 A23, November 2010.
\newblock \doi{10.1051/0004-6361/201015329}.

\bibitem[{von Paris} et~al.(2011){von Paris}, {Cabrera}, {Godolt}, {Grenfell},
  {Hedelt}, {Rauer}, {Schreier}, and {Stracke}]{vparis2011}
P.~{von Paris}, J.~{Cabrera}, M.~{Godolt}, J.~L. {Grenfell}, P.~{Hedelt},
  H.~{Rauer}, F.~{Schreier}, and B.~{Stracke}.
\newblock {Spectroscopic characterization of the atmospheres of potentially
  habitable planets: GL 581 d as a model case study}.
\newblock \emph{\aa}, 534:\penalty0 A26, October 2011.
\newblock \doi{10.1051/0004-6361/201117091}.

\bibitem[{Walker} et~al.(1981){Walker}, {Hays}, and {Kasting}]{walker1981}
J.~C.~G. {Walker}, P.~B. {Hays}, and J.~F. {Kasting}.
\newblock {A negative feedback mechanism for the long-term stabilization of the
  earth's surface temperature}.
\newblock \emph{\jgr}, 86:\penalty0 9776--9782, October 1981.
\newblock \doi{10.1029/JC086iC10p09776}.

\bibitem[{Wiscombe} and {Evans}(1977)]{wiscombe1977}
W.~J. {Wiscombe} and J.~{Evans}.
\newblock {Exponential-sum fitting of radiative transmission functions.}
\newblock \emph{J. Computational Physics}, 24:\penalty0 416--444, 1977.

\bibitem[{Wordsworth} et~al.(2011){Wordsworth}, {Forget}, {Selsis}, {Millour},
  {Charnay}, and {Madeleine}]{wordsworth2011}
R.~D. {Wordsworth}, F.~{Forget}, F.~{Selsis}, E.~{Millour}, B.~{Charnay}, and
  J.-B. {Madeleine}.
\newblock {Gliese 581d is the First Discovered Terrestrial-mass Exoplanet in
  the Habitable Zone}.
\newblock \emph{\apjl}, 733:\penalty0 L48, June 2011.
\newblock \doi{10.1088/2041-8205/733/2/L48}.

\end{thebibliography}

\end{document}